\documentclass[a4paper,11pt]{article}

\usepackage{jcappub}
\usepackage{tensor}     
\usepackage{graphicx}   
\usepackage[
colorlinks=true,        
citecolor=blue,         
linkcolor=blue,         
urlcolor=blue          
]{hyperref}             
\usepackage{bm}         
\usepackage{xcolor}     
\newcommand{\newc}{\newcommand*}
\newc{\figurewidth}{3.2in}
\newc{\la}{\lambda}
\newc{\La}{\Lambda}
\newc{\io}{\iota}
\newc{\Msun}{M_\odot}
\newc{\lvc}{LIGO-Virgo}
\newc{\PBH}{\mathrm{PBH}}
\newc{\ABH}{\mathrm{ABH}}

\def\({\left(}
\def\){\right)}
\def\[{\left[}
\def\]{\right]}

\def\ee{\begin{equation}}
\def\q{\end{equation}}
\def\m{\begin{eqnarray}}
\def\n{\end{eqnarray}}
\def\a{\begin{aligned}}
\def\b{\end{aligned}}

\newc{\red}[1]{\textcolor{red}{#1}}
\newc{\yellow}[1]{\textcolor{yellow}{#1}}
\newc{\green}[1]{\textcolor{green}{#1}}
\newc{\blue}[1]{\textcolor{blue}{#1}}
\begin{document}
\title{Merger rate of charged black holes from the two-body dynamical capture}
\author[a]{Lang Liu,}
\author[a]{Sang Pyo Kim}

\affiliation[a]{Department of Physics, Kunsan National University, Kunsan 54150, Korea}

\emailAdd{liulang915871@gmail.com}
\emailAdd{sangkim@kunsan.ac.kr}

\date{\today}
\abstract{We consider the two-body dynamical capture of black holes carrying U(1) charge which can not only correspond to electric or magnetic charge but also have other physical interpretations such as dark or hidden charge.  In the low-velocity and weak-field regime, we study gravitational and electromagnetic radiations from point masses with U(1) charges in a hyperbolic orbit, and we develop a formalism to derive the merger rate of charged black holes from the two-body dynamical capture.  We apply the formalism to find the effects of the charge-to-mass ratio on the merger rate for possible different cases and discover that the effects depend on the models.}

\maketitle
\section{Introduction}\label{intro}

Nearly one hundred years after the theoretical prediction of gravitational waves (GWs) in the general relativity (GR) by Einstein~\cite{Einstein:1916cc, Einstein:1918btx}, the first direct measurement of GWs \cite{Abbott:2016blz} by the Advanced Laser Interferometer Gravitational-Wave Observatory (LIGO) detectors established gravitational wave astronomy.  The LIGO-Virgo detection opened a new window to observe the universe and marked the dawn of multi-messenger astronomy \cite{LIGOScientific:2018mvr, Abbott:2020niy, LIGOScientific:2021djp}. So far, the LIGO-Virgo detection results have already provided extremely accurate confirmation that all gravitational merger events can be described by GR \cite{LIGOScientific:2019fpa, LIGOScientific:2020tif, LIGOScientific:2021sio}.

The no-hair theorem of black holes (BHs) in GR states that four-dimensional stationary BHs in the Einstein-Maxwell theory can be completely described by a Kerr-Newman metric~\cite{Hawking:1971vc, Robinson:1975bv, Cardoso:2016ryw}, and the theorem allows one to characterize all BH solutions in asymptotically flat spacetime by three physical quantities: the mass, spin and charge. When Hawking radiation is not considered,
in comparison with Schwarzschild BHs, charged BHs emit not only gravitational radiation but also electromagnetic radiation, and have attracted much attention \cite{Zilhao:2012gp,Zilhao:2013nda,Liebling:2016orx,Toshmatov:2018tyo,Bai:2019zcd,Allahyari:2019jqz,Christiansen:2020pnv,Wang:2020fra,Bozzola:2020mjx,Kim:2020bhg,Cardoso:2020nst, McInnes:2020gxx,Bai:2020ezy,Diamond:2021scl,Bozzola:2021elc,McInnes:2021frb,Kritos:2021nsf,Hou:2021suj,Benavides-Gallego:2021the,Diamond:2021dth}. Recently, assuming that the influence of BH spins can be neglected, Bozzola et al found that the charge-to-mass ratios of up to 0.3 are compatible with Binary BH merger event GW150914 \cite{Bozzola:2020mjx}. Later, they also reported general relativistic simulations of the inspiral and merger of non-spinning charged binary BHs \cite{Bozzola:2021elc}.

In the universe, the two-body dynamical capture is a fairly common and effective way to form the binary BH systems including astrophysical black holes (ABHs) and primordial black holes (PBHs)~\cite{Mouri:2002mc, Bird:2016dcv}. Therefore, it is important and meaningful to work out the merger rate of charged BHs from the two-body dynamical capture. To do so, we analytically find gravitational and electromagnetic radiations from a binary of point masses with charges in a hyperbolic orbit, and calculate the merger rate of BHs from the two-body dynamical capture with charges and a general mass function by taking into account gravitational and electromagnetic radiation.
The U(1) charge considered in this paper can correspond to the following physical interpretations: (1) electric charges, (2) magnetic charges \cite{Maldacena:2020skw,Bai:2020spd,Ghosh:2020tdu}, (3) hidden or dark  charges interacting with dark electromagnetism \cite{Ackerman:2008kmp,Feng:2009mn,Foot:2014uba,Foot:2014osa}, (4) modified theories of gravity with additional scalar or vector fields \cite{Moffat:2005si,Cardoso:2016olt,Cardoso:2020iji}, and (5) the fifth force \cite{Fischbach:1992fa,Dvali:2001dd,Gubser:2004uf}.

The paper is organized as follows. In Section \ref{II}, in the low-velocity and weak-field regime, we calculate gravitational and electromagnetic radiations from point masses with U(1) charges in a hyperbolic orbit. In Section \ref{III}, we develop a formalism to derive the merger rate of charged BHs from the two-body dynamical capture via gravitational and electromagnetic radiations. In Section \ref{IV}, we show the effects of the charge-to-mass ratio on the merger rate of ABH and PBH binaries from the two-body dynamical capture. Finally, we summarize our results and conclude with physical implications in the last section.

In this paper, we set $G=c =4 \pi \varepsilon_{0} = \frac{\mu_0}{4\pi}=1$ unless otherwise specified. Although the U(1) charges are intended to have different physical interpretations, we will refer to their quantities by using ``electromagnetic" for the sake of simplification through this paper.

\section{Gravitational and electromagnetic radiations from point masses in a hyperbolic orbit}
\label{II}

\begin{figure}[htbp!]
\includegraphics[width=0.8\textwidth]{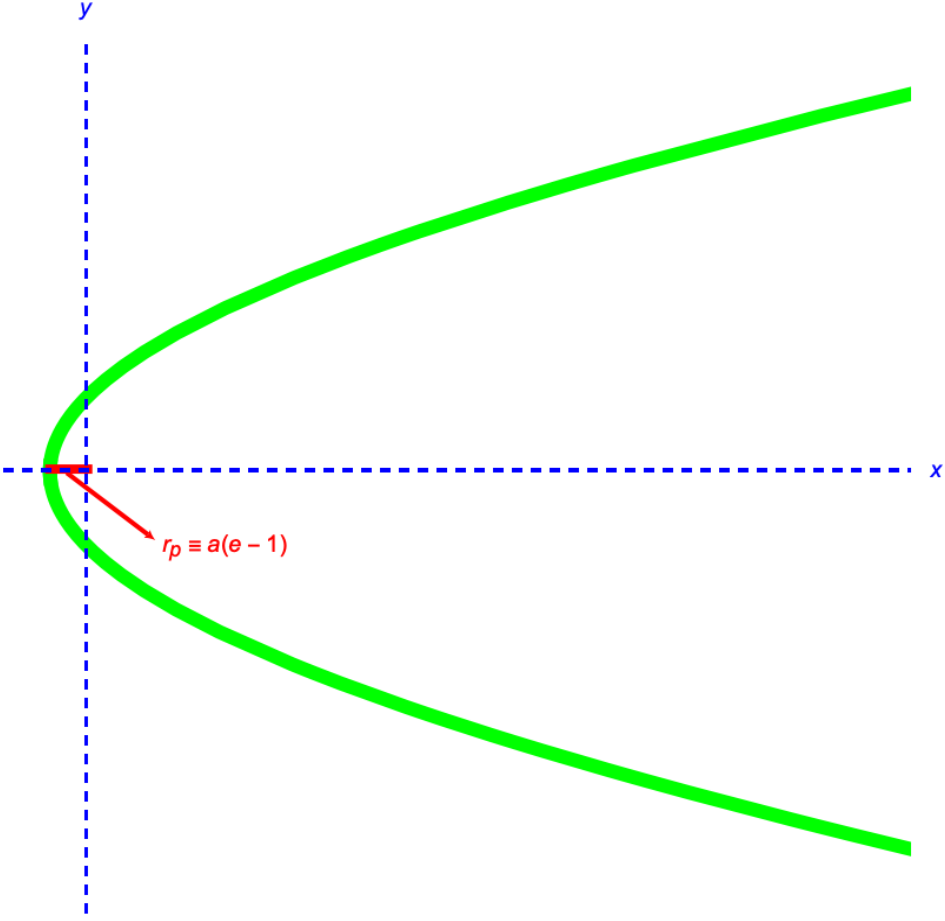}
\caption{\label{fig:a}
A schematic picture of a hyperbolic orbit.}
\end{figure}

{ We study an encounter of two unbounded massive charges and the gravitational and electromagnetic radiations from the encounter.
The unbound system under gravitational and electric forces follows a hyperbolic orbit and dominantly emits gravitational and electromagnetic radiations near the closest approach and loses energy and angular momentum.
To describe the dynamical evolution, } we assume that the orbit lies in $x$-$y$ plane, and that the coordinates of the point masses $m_1$ with charge $Q_1$ and $m_2$ with charge $Q_2$ are ($d_1\cos{\psi}$, $d_1 \sin{\psi}$) and ($-d_2 \cos{\psi}$, $-d_2\sin{\psi}$), respectively. Choosing the origin at the center of mass, we have
\ee
\label{d1d2}
d_{1}=\left(\frac{m_{2}}{m_{1}+m_{2}}\right) d, \quad d_{2}=\left(\frac{m_{1}}{m_{1}+m_{2}}\right) d.
\q
As shown in Fig.~\ref{fig:a}, the equation for a hyperbolic orbit is
\ee
\label{d}
d=\frac{a\left(e^{2}-1\right)}{1-e \cos \psi},
\q
where $a$ and $e$, which can be interpreted as the semi-major axis and eccentricity, and are defined as
\ee
a\equiv \frac{m_1 m_2(1-\la)}{2E},~\quad
e\equiv \left(1+\frac{2 E {L}^{2}}{\left(m_1+m_2\right)m_1 m_2 \left(1-\la\right)^2}\right)^{1 / 2}.
\q
Here, $\la \equiv \frac{Q_1Q_2}{m_1m_2}$ represents the ratio of the Coulomb force to the gravitational force,  $E$ is the total energy of charged binary system including the gravitational, electrostatic and kinetic energy, and $L$ is the angular momentum of binary system. And the angular velocity along the orbit is given by
\ee
\label{psi}
\dot{\psi}=\frac{\left[\left(m_{1}+m_{2}\right) a\left(e^{2}-1\right) (1-\la)\right]^{1 / 2}}{d^{2}}.
\q

Firstly, we calculate the gravitational radiation from the point masses $m_1$ with charge $Q_1$ and $m_2$ with charge $Q_2$ in the hyperbolic orbit.
The non-vanishing second mass moment is a $2 \times 2$ matrix that is given by
\begin{equation}
\label{Mab}
M_{a b}=\mu d^{2}\left(\begin{array}{cc}{\cos ^{2} \psi} & {\sin \psi \cos \psi} \\ {\sin \psi \cos \psi} & {\sin ^{2} \psi}\end{array}\right)_{a b},
\end{equation}
where subscripts $\(a, b=1,2\)$ refer to indices in the $x$-$y$ plane and $\mu=m_1m_2/(m_1+m_2)$ is the reduced mass. Note that
$M_{ij}$ is reducible while the traceless quadrupole moment $Q_{ij} \equiv M_{ij}-\frac{1}{3} \delta_{ij} M_{kk}$ is irreducible.
Following \cite{Peters:1963ux}, the rate of energy loss from a system by  gravitational quadrupole radiation is given by
\ee
\a
\label{dEGWdt}
\frac{dE_{GW}^{quad}}{dt} \equiv  -\frac{1}{5}\left(\dddot{Q}_{ij}\dddot{Q}_{ij}\right)=-\frac{2}{15 }\left[\bigl( \dddot{M}_{11} + \dddot{M}_{22} \bigr)^2 - 3 \bigl( \dddot{M}_{11} \dddot{M}_{22} -  \dddot{M}_{12}^{2} \bigr) \right].
\b
\q
The derivatives of the components of the second mass moment tensor can be calculated by the aid of Eqs.~\eqref{d} and \eqref{psi}; putting those expressions into Eq.~\eqref{dEGWdt}, we obtain
\ee
\a
\label{dEGW}
\frac{dE_{GW}^{quad}}{dt}=-
\frac{4(1-\la)^3 m_1^2 m_2^2 \left(m_1+m_2\right) (e \cos (\psi )-1)^4 \left(11 e^2 \cos (2 \psi )+13 e^2-48 e \cos (\psi )+24\right)}{15 a^5 \left(e^2-1\right)^5}.
\b
\q
The post-Newtonian formalism gives the rate of angular momentum emission due to gravitational quadrupole radiation as ~\cite{Peters:1964zz}
\ee
\frac{d L_{GW}^{i,quad}}{d t}\equiv -\frac{2}{5 } \epsilon^{i k l}\left(\ddot{Q}_{k a} \dddot{Q}_{l a}\right) =-\frac{2}{5 } \epsilon^{i k l}\left(\ddot{M}_{k a} \dddot{M}_{l a}\right),
\q
where $\epsilon^{i k l}$ is the Levi-Civita symbol ($\epsilon^{1 2 3} =1$). As the orbit is assumed to lie in the $x$-$y$ plane and the gravitational and Coulomb forces keep the same orbital plane, we have $L^z=L$ and $L^x=L^y=0$. Substituting Eqs.~\eqref{d} and \eqref{psi} into the components of the second mass moment tensor for the unbound system in the hyperbolic orbit, the non-vanishing rate of angular momentum emission due to gravitational quadrupole radiation is
\ee
\a
\label{dLGW}
\frac{d L_{GW}^{quad}}{d t}&=\frac{d L_{GW}^{z,quad}}{d t}=-\frac{8  (1-\la)^{5/2}m_1^2 m_2^2 \sqrt{m_1+m_2}}{5 a^{7/2} \left(e^2-1\right)^{7/2}} \sin ^2(\psi ) (e \cos (\psi )-1)^2
 \\
&\times (e (\cos (2 \psi )+3)-4 \cos (\psi )) (e (3 \cos (2 \psi )+4)-8 \cos (\psi )).
\b
\q

Secondly, we compute the energy and angular momentum loss rates due to electromagnetic dipole radiation. In our reference frame, the electric dipole is given by
\ee
\label{PQ}
\bm{p}\equiv Q_1 \bm{d_1}+Q_2 \bm{d_2}=\frac{m_2Q_1-m_1Q_2}{m_1+m_2} (d \cos \psi \hat{\bm{x}}+ d \sin \psi \hat{\bm{y}}).
\q
{The overall factor can be written as $\mu (\frac{Q_1}{m_1} - \frac{Q_2}{m_2})$.} Then, according to \cite{Landau:1982dva,Liu:2020cds}, the energy and angular momentum loss rates due to electromagnetic dipole radiation are
\ee
\label{dEEM}
\frac{dE_{EM}^{dip}}{dt}=-\frac{2\ddot{p}^2}{3}=-\frac{2  (1-\la)^2 \left(m_2 Q_1-m_1 Q_2\right){}^2 (e \cos (\psi )-1)^4}{3 a^4 \left(e^2-1\right)^4},
\q
\ee
\label{dLEM}
\frac{d L_{EM}^{dip}}{d t}= -\frac{2}{3} (\dot{p}_2 \ddot{p}_1-\dot{p}_1 \ddot{p}_2)=-\frac{2(1-\la)^{3/2} \left(m_2 Q_1-m_1 Q_2\right){}^2 (1-e \cos (\psi ))^3}{3 a^{5/2} \left(e^2-1\right)^{5/2} \sqrt{m_1+m_2}}.
\q
In Appendix \ref{appendix}, we calculate the quadrupole term of electromagnetic radiation and find that its contribution is always smaller than the quadrupole term of gravitational radiation. The total energy and angular momentum radiated due to gravitational quadrupole radiation and electromagnetic dipole radiation are then the integral of Eqs.~\eqref{dEGW}, \eqref{dLGW}, \eqref{dEEM} and \eqref{dLEM} over the orbit.  Notice the asymptotes of the hyperbolic orbit are $y=\pm \tan(\psi_0) x$, where $\psi_0$ is determined by $\cos(\psi_0)=1/e$, i.e, $\psi \in (\psi_0, 2\pi-\psi_0)$; thus we find
\ee
\a
\label{a1}
\Delta E_{GW}^{quad}&=\int^{\infty}_{-\infty}\frac{dE_{GW}^{quad}}{dt} dt=\int^{2\pi-\psi_0}_{\psi_0}\frac{dE_{GW}^{quad}}{dt}(\frac{d \psi}{dt})^{-1} d\psi=-\frac{2(1-\la)^{5/2} m_1^2 m_2^2 \sqrt{m_1+m_2}}{45 a^{7/2} \left(e^2-1\right)^{7/2}}
\\
& \times \left(3 \left(37 e^4+292 e^2+96\right) \left(\pi -\cos ^{-1}\left(\frac{1}{e}\right)\right) \pi +\sqrt{1-\frac{1}{e^2}} e \left(673 e^2+602\right) \right),
\b
\q
\ee
\a
\label{a2}
\Delta L_{GW}^{quad}&=\int^{\infty}_{-\infty}\frac{dL_{GW}^{quad}}{dt} dt=\int^{2\pi-\psi_0}_{\psi_0}\frac{dL_{GW}^{quad}}{dt}(\frac{d \psi}{dt})^{-1} d\psi=\frac{8(1-\la)^2 m_1^2 m_2^2}{5 a^2 \left(e^2-1\right)^2}
\\
& \times \left(\left(7 e^2+8\right) \left(\pi -\cos ^{-1}\left(\frac{1}{e}\right)\right)+\frac{\sqrt{1-\frac{1}{e^2}} \left(21 e^4-10 e^2+4\right)}{e^3}\right),
\b
\q
\ee
\a
\label{a3}
\Delta E_{EM}^{dip}&=\int^{\infty}_{-\infty}\frac{dE_{EM}^{dip}}{dt} dt=\int^{2\pi-\psi_0}_{\psi_0}\frac{dE_{EM}^{dip}}{dt}(\frac{d \psi}{dt})^{-1} d\psi=-\frac{(1-\la)^{3/2} \left(m_2 Q_1-m_1 Q_2\right){}^2}{3 a^{5/2} \left(e^2-1\right)^{5/2} \sqrt{m_1+m_2}}
\\
& \times \left(\pi  \left(e^2+2\right)+6 e \sqrt{1-\frac{1}{e^2}}+2 \left(e^2+2\right) \csc ^{-1}(e)\right),
\b
\q
\ee
\a
\label{a4}
\Delta L_{EM}^{dip}&=\int^{\infty}_{-\infty}\frac{dL_{EM}^{dip}}{dt} dt=\int^{2\pi-\psi_0}_{\psi_0}\frac{dL_{EM}^{dip}}{dt}(\frac{d \psi}{dt})^{-1} d\psi=-\frac{4(1-\la) \left(m_2 Q_1-m_1 Q_2\right){}^2}{3 a \left(e^2-1\right) \left(m_1+m_2\right)}
\\
& \times \left(e \sqrt{1-\frac{1}{e^2}}-\cos ^{-1}\left(\frac{1}{e}\right)+\pi \right).
\b
\q
{In the case of either charge neutral masses ($Q_1=Q_2=\la=0$) or the same charge to mass ratio ($\frac{Q_1}{m_1} = \frac{Q_2}{m_2}$), } we find that the electric dipole radiation vanishes, $\Delta E_{EM}^{dip}=\Delta L_{EM}^{dip}=0$, and then the total energy and angular momentum are radiated through the gravitational quadrupole radiation and agree with the results given by Hansen \cite{Hansen:1972jt}. In the parabolic limit, i.e., $e\rightarrow 1$, Eqs.~\eqref{a1}, \eqref{a2}, \eqref{a3} and \eqref{a4} can be simplified as
\ee
\label{b1}
\Delta E_{EM}^{dip}=-\frac{\pi (1-\la)^{3/2} \left(m_2 Q_1-m_1 Q_2\right){}^2}{2 \sqrt{2} \sqrt{m_1+m_2} r_p^{5/2}},
\q
\ee
\label{b2}
\Delta E_{GW}^{quad}=-\frac{85 \pi  (1-\la)^{5/2} m_1^2 m_2^2 \sqrt{m_1+m_2}}{12 \sqrt{2} r_p^{7/2}},
\q
\ee
\label{b3}
\Delta L_{EM}^{dip}=-\frac{2 \pi(1-\la) \left(m_2 Q_1-m_1 Q_2\right){}^2}{3 \left(m_1+m_2\right) r_p},
\q
\ee
\label{b4}
\Delta L_{GW}^{quad}=-\frac{6 \pi (1-\la)^2 m_1^2 m_2^2}{r_p^2},
\q
where $r_p \equiv a(e-1)$ is the distance of periastron. Another way to estimate the total energy loss is to approximate the trajectory of a close encounter by taking the $e=1$ limit of the elliptical orbit {since the radiation is most dominant when the orbit is closest to the focus}; thereby the total energy loss is given by $\Delta E=T\left\langle\frac{d E}{d t}\right\rangle$ where $T$ is orbital period and $\left\langle\frac{d E}{d t}\right\rangle$ is the time-averaged energy loss rate of the binary in the Keplerian orbit which is given by \cite{Liu:2020cds}. The gravitational and electromagnetic radiations from binary BHs with electric and magnetic charges were found in \cite{Liu:2020vsy, Liu:2020bag, Liu:2022cuj}.

\section{Merger rate of charged black holes from the two-body dynamical capture}
\label{III}

If two charged BHs get closer and closer, the energy loss due to gravitational and electromagnetic radiations can exceed the orbital kinetic energy, and thus the unbound system cannot escape to infinity anymore and form a binary with a negative orbital energy. This binary immediately merges through consequent gravitational and electromagnetic radiations\cite{Liu:2020cds}. From such a process, we can estimate the cross section and calculate the rate coefficient.

Let us consider the interaction of two charged BHs with masses $m_1$, $m_2$ and charges $Q_1$, $Q_2$ and assume that they have an initial relative velocity $v$, the distance of periastron $r_p $ and the impact parameter $b$. We can approximate the trajectory of a close encounter by the hyperbolic with $e\rightarrow 1$ since when the two charged BHs pass by closely, the true trajectory is physically indistinguishable from a parabolic one near the periastron, in which gravitational and electromagnetic radiations dominantly occur. According to Sec. \ref{II} and Appendix \ref{appendix}, the total energy loss due to the electromagnetic dipole and quadrupole radiations and the gravitational quadrupole radiation by the close-encounter can be evaluated by using $e \rightarrow 1$ and the definition of $r_p \equiv a(e-1)$,
\ee
\a
\label{dE}
\Delta E&=\Delta E_{EM}^{dip}+\left(1+ \Lambda \right) \Delta E_{GW}^{quad},
\b
\q
where we denote $\Lambda=\frac{\mu^{2}\left(Q_{1} / m^2_{1}+Q_{2}/ m^2_{2}\right)^{2}}{4}$ for short and $\Delta E_{EM}^{dip}$, $\Delta E_{GW}^{quad}$ are given by Eqs.~\eqref{b1} and \eqref{b2}. Similarly, the total angular momentum loss due to the electromagnetic dipole and quadrupole radiations and the gravitational quadrupole radiation is given by
\ee
\Delta L=\Delta L_{EM}^{dip}+\left(1+ \frac{\Lambda}{2} \right) \Delta L_{GW}^{quad},
\q
where $\Delta L_{EM}^{dip}$, $\Delta L_{GW}^{quad}$ are given by Eqs.~\eqref{b3} and \eqref{b4}. The relation between $r_p$ and $b$ is $b^{2}=r_{p}^{2}+\frac{2 \left(1-\la\right) (m_i+m_j) r_{p}}{v^{2}}$. In the limit of a strong gravitational and electromagnetic focusing (i.e. $r_p\ll b$), then the distance of closest approach $r_p$ is given by
\ee
\label{rp}
r_{p }=\frac{b^{2} v^{2}}{2\left(1-\la \right)\left(m_{i}+m_{j}\right)}.
\q
The condition for the two charged BHs to form a binary is that the total energy loss due to radiations is larger than the kinetic energy $\mu v^2/2$, i.e.,
\ee
\label{Condition}
\Delta E+\frac{\mu v^2}{2}<0.
\q
Equations \eqref{dE},~\eqref{rp} and \eqref{Condition} yield the maximum impact parameter $b_{max}$ for the charged BHs to form a bound system, and then we obtain the merging cross section as $\sigma=\pi b_{max}^2$, where $b_{max}$ is determined by
\ee
\label{b}
\frac{2 \pi (1-\la)^4 \left(m_1+m_2\right){}^2 \left(m_2 Q_1-m_1 Q_2\right)^2}{b_{max}^5 v^5}+\frac{170 \pi (1-\la)^6 \Lambda  m_1^2 m_2^2 \left(m_1+m_2\right)^4}{3 b_{max}^7 v^7}=\frac{\mu v^2}{2}.
\q
If the dipole radiation is dominant (i.e. $\Delta E_{EM}^{dip}> (1+\La) \Delta E_{GW}^{quad} $ ), the merging cross section is
\ee
\label{dipole}
\sigma_{dip}\approx\frac{2^{4/5} \pi ^{7/5} (1-\la)^{8/5} \left(m_1+m_2\right){}^{6/5} \left(m_1 Q_2-m_2 Q_1\right){}^{4/5}}{m_1^{2/5} m_2^{2/5} v^{14/5}},
\q
whereas if the total quadrupole radiation is dominant (i.e. $\Delta E_{EM}^{dip}< (1+\La) \Delta E_{GW}^{quad} $ ), the merging cross section is
\ee
\a
\label{quad}
\sigma_{quad}\approx\frac{\left(\frac{85}{3}\right)^{2/7} 2^{4/7} \pi ^{9/7} (1-\la)^{12/7} m_1^{2/7} m_2^{2/7} \left(m_1+m_2\right){}^{10/7}}{v^{18/7}} \left(1+ \Lambda \right)^{2/7},
\b
\q
which is consistent with \cite{Mouri:2002mc}.
Notice that the Schwarzschild radius $\sigma_{Sch} \simeq \pi (2G m)^2 \simeq (v)^{18/7} \sigma_{GW}$, $v \ll 1$ and $\sigma \geq \sigma_{GW}$; thereby the Newtonian approximation is sufficiently accurate no matter whether electromagnetic radiation is dominant or gravitational radiation is dominant.
The differential merger rate of charged BHs per comoving volume reads
\ee
\label{merger rate}
dR=n(m_1,Q_1)n(m_2,Q_2)\left\langle\sigma v\right\rangle dm_1dm_2dQ_1dQ_2
\q
where $n(m_1,Q_1)$ and $n(m_2,Q_2)$ are the comoving average number density of charged BHs with masses $m_1$, $m_2$ and charges $Q_1$, $Q_2$, and $\left\langle\sigma v\right\rangle$ denotes the average over relative velocity distribution with $\sigma=\pi b_{max}^2$ given in Eq.~\eqref{b}. In this section, we have computed the merging cross section of charged BHs without assuming the origin of those charged BHs.

\section{Effects of charge-to-mass ratio on the merger rate}
\label{IV}

\subsection{Astrophysical black hole}
In astrophysics, there are two main astrophysical
mechanisms for ABH binary formation\footnote{ There are others channels like mergers in triple systems assisted by the Kozai-Lidov mechanism (See a recent review \cite{Barack:2018yly} for details ). }. One channel is isolated (or ‘field’) binary evolution and the other is dynamical capture in dense stellar environments. Dynamical capture can become absolutely effective in deeply dense stellar environments. Especially, it is shown in Ref.~\cite{Wong:2020ise} that the vast majority ($\sim 80\%$) of the merger events detected by LIGO-Virgo comes from dynamical capture\footnote{There are currently some disagreements about the contributions of different channels. It is argued that in Ref. \cite{Zevin:2020gbd} that $\sim 90\%$ is provided by isolated binary evolution channel.}.

According to the initial mass function that describes the number distribution of stars \cite{Kroupa:2000iv}, we assume the mass function of ABHs takes a truncated power-law form as \cite{LIGOScientific:2020kqk}
\ee
p(m) \propto m^{\zeta} \theta\left(m-m_{\min }\right) \theta\left(m_{\max }-m\right)
\q
with $\zeta=-1.3$, and the normalization of mass function $\int p(m) d m=1$ will be taken. Here, we adopt $m_{\min }=3.0 M_{\odot}$ which corresponds to the lower mass bound of ABHs \cite{Rhoades:1974fn}, and  $m_{\max }=55 M_{\odot}$ which corresponds to the beginning of pair-instability supernova (PISN) gap \cite{Woosley:2002zz}.
We consider the following two simple but different models
\begin{itemize}
\item model (a): all ABHs have the  same charge-to-mass ratio $\iota$ in magnitude and sign,
\item model (b): a half of ABHs have the same charge-to-mass ratio $\iota$  and the other half ABHs have a charge-to-mass ratio equal in magnitude but oppositely charged.
\end{itemize}
In the following calculation, the Maxwell-Boltzmann distribution $P(v) \propto v^{2} \exp \left(-v^{2} / v_{0}^{2}\right)$ will be used for the velocity distribution of BHs with the most probable velocity $v_{0}=100$ km/s.

To show the effects of charge-to-mass ratio $\iota$ on the merger rate of ABH binaries from the two-body dynamical capture, we define a function of $\iota$ as
\ee
\label{eta}
\eta(\iota)\equiv\frac{R_{\ABH}(\iota)}{R_{\ABH}(0)},
\q
where $R_{\ABH}(\iota)$ is the total merger rate of charged ABHs with charge-to-mass ratio $\iota$ and $R_{\ABH}(0)$ is the total merger rate of ABHs in charge neutral case.

In the model a, we have the number density of charged ABHs
\ee
n(m,Q)\propto \delta(\frac{Q}{m}-\iota)\frac{p(m)}{m},
\q
where $\delta$ is the Dirac delta function. In such a case, we have $Q_{1}/m_{1}=Q_{2}/m_2=\iota$. Thus, the dipole radiation vanish and the energy loss due to total quadrupole radiation is given by $\Delta E=\(1+\io^2/4\) \Delta E_{GW}^{quad}$.  Therefore, according to Eqs.~\eqref{quad} and \eqref{merger rate} and from the definition of $\eta$, we obtain
\ee
\eta(\iota)=(1-\iota^2)^{12/7} \(1+\io^2/4\)^{2/7}.
\q

\begin{figure}[htbp!]
\centering
\includegraphics[width = 0.48\textwidth]{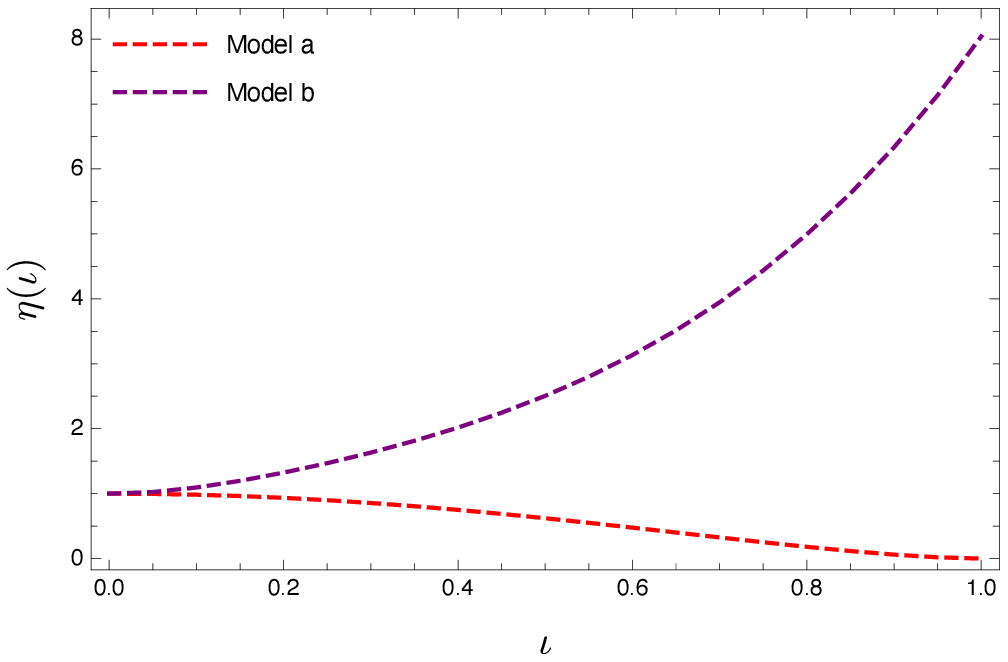}
\includegraphics[width = 0.48\textwidth]{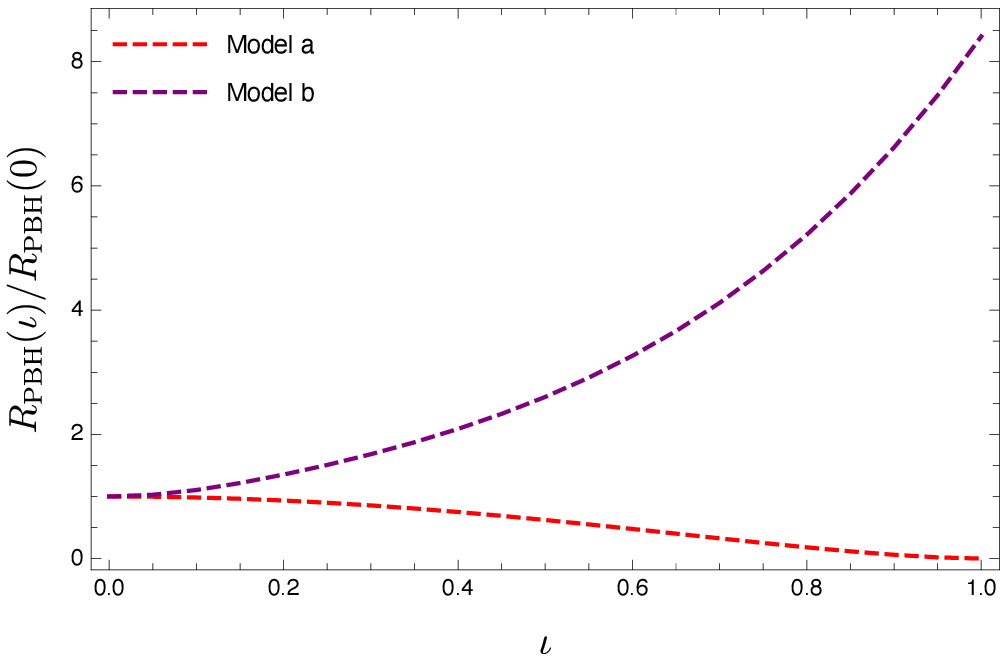}
\caption{\label{fig:1}
Left: The plots of $\eta(\io)$ as the function of $\iota$ in the model a (red) and model b (purple). Right: The normalized merger rate of charged PBH $R_{\PBH}(\io)/R_{\PBH}(0)$ as the function of $\iota$ in the model a (red) and model b (purple).}
\end{figure}

In the model b, we have the number density of charged ABHs
\ee
n(m,Q)\propto \frac{1}{2}\left(\delta(\frac{Q}{m}-\iota)+\delta(\frac{Q}{m}+\iota)\right) \frac{p(m)}{m}.
\q
Therefore, the merger events can be divided into two parts. One half of merger events are those binary BHs with the same sign charges while the other half of merger events are those binary BHs with the opposite charges. When the binary BHs have the opposite charges, we have  $Q_{1}/m_{1}=-Q_{2}/m_2=\iota$ and $\Lambda=\left(1+\frac{\iota ^2 \left(m_1-m_2\right){}^2}{4 \left(m_1+m_2\right){}^2}\right)$. The merging cross section in this case is  $\sigma \simeq \pi b_{max}^2  $, where $b_{max}$ is a root of
\ee
\frac{8 \pi  G^4 \iota ^2 \left(\iota ^2+1\right)^4 m_1^2 m_2^2 \left(m_1+m_2\right)^2}{b_{max}^5 v^5}+\frac{170 \pi  G^7 (1+\iota ^2)^6 \Lambda  m_1^2 m_2^2 \left(m_1+m_2\right)^4}{3 b_{max}^7 v^7}=\frac{\mu v^2}{2}.
\q
Hence, in the model b, from the Eqs.~\eqref{merger rate} and \eqref{eta}, we have
\ee
\eta(\io)=\frac{1}{2}(1-\iota^2)^{12/7} \(1+\io^2/4\)^{2/7}+\frac{1}{2}\frac{\int \int \int P(v)\sigma v\frac{p(m_1)}{m_1}\frac{p(m_2)}{m_2}dvdm_1dm_2}{\int \int \int P(v)\sigma_0 v\frac{p(m_1)}{m_1}\frac{p(m_2)}{m_2}dvdm_1dm_2}
\q
where
\ee
\sigma_0=\frac{2^{4/7} \left(\frac{85}{3}\right)^{2/7} \pi ^{9/7} m_1^{2/7} m_2^{2/7} \left(m_1+m_2\right){}^{10/7}}{v^{18/7}}
\q
represents the merging cross section for charge neutral BHs. From the definition above, we have $\eta(\io)=\eta(-\io)$ and $\eta(0)=1$ for either the model a or b. Therefore, we only need to consider $\iota \in [0,1]$. In Fig.~\ref{fig:1}, we plot $\eta(\io)$ as the function of $\iota$ for two models. In the model a, we find that $\eta(\io)$ decreases as $\iota$ increases and $\eta(1)=0$. For the model b, we exactly show that $\eta(\io)$ increases as $\iota$ increases and reaches the maximum value of $\eta(1) \approx 8.0$.

\subsection{Primordial black hole}

In contrast with ABHs, PBHs are those BHs that are formed in the radiation-dominated era of the early universe due to the collapse of large energy density fluctuations \cite{Zeldovich:1967lct, Hawking:1971ei, Carr:1974nx}.
The issue of how to distinguish ABHs and PBHs has been studied in \cite{Chen:2019irf, Mukherjee:2021ags, Liu:2021jnw, Franciolini:2021xbq}.
In recent years, PBHs have attracted a lot of attention and have been extensively studied, not only because they can be candidates of dark matter~\cite{Carr:2016drx}, but also because they can be the sources for \lvc~detections~\cite{Bird:2016dcv, Sasaki:2016jop} and the seeds for galaxy formation~\cite{Bean:2002kx, Kawasaki:2012kn, Carr:2018rid}. There are two primary mechanisms for PBH binary formation, which can be classified by the cosmological epoch when they work. One mechanism is that the PBH binary would have formed in the late universe \cite{Bird:2016dcv,Clesse:2016vqa,Nishikawa:2017chy} while the other is that PBH binary would have formed in the early universe \cite{Nakamura:1997sm,Ioka:1998nz,Sasaki:2016jop,Raidal:2017mfl,Kocsis:2017yty,Ali-Haimoud:2017rtz,Chen:2018czv,Ballesteros:2018swv,Raidal:2018bbj,Liu:2018ess,Liu:2019rnx,Young:2019gfc,Vaskonen:2019jpv,Garriga:2019vqu}. Recently, it is shown in Ref~\cite{Kritos:2021nsf} that PBHs can be near-extremally charged. In this subsection, we will investigate the effects of charge-to-mass ratio $\iota$ on the merger rate of PBH binaries which would have formed in the late universe. Similarly to ABHs, we also consider two special models by assuming that all PBHs have the same mass $M_{\PBH}$:
\begin{itemize}
\item model (a): all PBHs have the same charge-to-mass ratio $\iota$ in magnitude and sign,

\item model (b): a half of PBHs have the same charge-to-mass ratio $\iota$  and the other half of PBHs have the same magnitude but with the opposite charges.
\end{itemize}

In the model a, the number density of charged PBHs is given by
\ee
n(m,Q)=\frac{f\rho_{\mathrm{DM}}}{M_{\PBH}} \delta(\frac{Q}{m}-\iota) \delta(m-M_{\PBH}),
\q
where $f$ is the fraction of PBHs in the dark matter and $\rho_{\mathrm{DM}}$ is the dark matter energy density in the present time. According to Sec.~\ref{III}, the merger rate of the charged PBHs that would have formed in the late universe is
\ee
R_{\PBH}(\io)\approx 1.5\times 10^{-8} f^{2}(1-\iota^2)^{12/7} \(1+\io^2/4\)^{2/7} \mathrm{Gpc}^{-3} \mathrm{yr}^{-1},
\q
and is independent of the mass of the charged PBHs. For the model b, the number density of charged PBHs is
\ee
n(m,Q)=\frac{1}{2}\frac{f\rho_{\mathrm{DM}}}{M_{\PBH}} \left(\delta(\frac{Q}{m}-\iota)+\delta(\frac{Q}{m}+\iota)\right) \delta(m-M_{\PBH}).
\q
Following Sec.~\ref{III} again, the merger rate of the charged PBHs that would have formed in the late universe can be simplified as
\ee
\label{RB}
R_{\PBH}(\io)\approx 0.75\times 10^{-8} f^{2}(1-\iota^2)^{12/7} \(1+\io^2/4\)^{2/7} \mathrm{Gpc}^{-3}\mathrm{yr}^{-1} +\frac{f^2\rho_{\mathrm{DM}}^2}{2M_{\PBH}^2} \int P(v)\sigma_1 v dv,
\q
where $\sigma_1=\pi b_{max}^2 $ denotes the merging cross section for charged PBHs and $b_{max}$ is determined by
\ee
\label{bb}
-\frac{2720 \pi  \left(\iota ^2+1\right)^6 M_{\PBH}^8}{3 b_{max}^7 v^7}-\frac{32 \pi  \iota ^2 \left(\iota ^2+1\right)^4 M_{\PBH}^6}{b_{max}^5 v^5}+\frac{M_{\PBH} v^2}{4}=0.
\q
Notice that from Eq.~\eqref{bb} it follows $b_{max}\propto M_{\PBH}$ and $\sigma_1 \propto M_{\PBH}^2$. Therefore, we find that $R_{\PBH}(\io)$ in~\eqref{RB} is independent of $M_{\PBH}$ and scales as $f^2$. So, $R_{\PBH}(\io)/R_{\PBH}(0)$ is only a function of $\io$ where $R_{\PBH}(0)\approx 1.5\times 10^{-8} f^{2} \mathrm{Gpc}^{-3} \mathrm{yr}^{-1}$.   In Fig.~\ref{fig:1}, we plot the normalized merger rate $R_{\PBH}(\io)/R_{\PBH}(0)$ of charged PBHs  as the function of $\iota$ for different models. Similarly, we notice that $\eta(\io)$ decreases as $\iota$ increases and $\eta(1)=0$ in the model a while $\eta(\io)$ increases as $\iota$ increases and reaches $\eta(1) \approx 8.4$ in the model b.

In this section, we have shown the effects of the charge-to-mass ratio on the merger rate of ABH and PBH binaries from the two-body dynamical capture for possible different models. Here, we want to discuss the understandings of the different models.  As mentioned in the introduction, the U(1) charge considered in this paper can correspond to the different physical models and have interesting interpretations. When the U(1) charge corresponds to modified theories of gravity with additional scalar or vector fields, we can choose model a as a representative model. On the other hand, when the U(1) charge corresponds to electric, magnetic or dark charge, we can choose model b as a representative model, which suggests that the part of the universe under study is charge neutral but charges are locally separated by some mechanisms which were found in Refs.~\cite{Wald:1974np, Bai:2019zcd}.

\section{Conclusion}\label{Con}

In this work, we have calculated gravitational and electromagnetic radiations from binaries of point masses with U(1) charges in a hyperbolic orbit in the low-velocity and weak-field regime, and applied the result to derive the merger rate of charged BHs from the two-body dynamical capture via gravitational and electromagnetic radiations. We have then shown the effects of the charge-to-mass ratio on the merger rate of ABH and PBH binaries from the two-body dynamical capture. We have also found that the effects of the charge-to-mass ratio on the merger rate depend on the models.

There are various possible extensions and applications of BHs with U(1) charges.  Firstly, the U(1) charges can be interpreted as electric charges. As shown in~\cite{zhang:2016rli, Liu:2016olx}, the mergers of electrically charged BHs may potentially offer an explanation for the still-mysterious fast radio bursts and gamma-ray bursts. In Refs.~\cite{Deng:2018wmy,Deng:2021pqa}, the coalescence of earth-mass primordial BHs with the charge-to-mass ratio $\sim 10^{-4}$ is proposed to be a candidate for sources of the observed fast radio bursts. Here, we have confirmed that the effect of the charge-to-mass ratio of $\sim 10^{-4}$ {is negligible and the merger scenarios of charged BHs are similar to those without charges.} Secondly, the U(1) charges can be interpreted as magnetic charges. Magnetic charges, if they would have existed in the early universe, will offer a new window to explore fundamental physics. Recently, Maldacena extensively discussed the spectacular electroweak symmetry restoration of magnetically charged BHs \cite{Maldacena:2020skw}. Since no evidence of magnetic charges or monopoles has been found yet from terrestrial experiments ~\cite {Staelens:2019gzt, Mavromatos:2020gwk}, the merger events of binary BHs detected by LIGO-Virgo may provide us with a new way to investigate whether these BHs indeed have magnetic charges. Finally, the U(1) charges can be interpreted as dark or hidden charges with an extremely weak coupling constant. It is shown in Refs~\cite{Wang:2020fra, Bozzola:2020mjx} that some merger events detected by LIGO-Virgo are compatible with charged BHs while some are compatible with uncharged BHs. An interesting suggestion to this controversy is that those merger events would have originated from different scenarios. Therefore, a key question is how those BHs would carry dark or hidden charges. We leave all those interesting issues for future works.

\acknowledgments
We would like to thank Daniel Marin for the useful conversations. This work was supported in part by the National Research Foundation of Korea (NRF) funded by the Ministry of Education (2019R1I1A3A01063183).


\appendix
\section{Appendix}
\subsection{Electromagnetic dipole and quadrupole radiations}
\label{appendix}
In the case of the same charge to mass ratio ($Q_1/m_1=Q_2/m_2$), the electromagnetic dipole radiation vanishes and the quadrupole radiation becomes the leading term. This is true for any system of charges~\cite{Landau:1982dva}. Here, we will calculate the energy and angular momentum loss rates due to the electromagnetic dipole and quadrupole radiations. From the definition of traceless charge quadrupole, we find
\ee
\label{definition}
D^{a b}=\sum_{i=1,2}Q_i d_i^a d_i^b-\frac{1}{3} \sum_{i=1,2}Q_i d_i^c d_i^c \delta_{ab}=\mu\left(\frac{Q_1}{m_1^2} +\frac{Q_1}{m_1^2}\right) Q^{a b}.
\q
We extend the magnetic vector potential $\bm{A}$ as the quadrupole term,
\ee
\label{A}
A_i=\frac{P_{ij}(\bm{n})}{r}(\dot{p}^j+\frac{1}{2} \ddot{D}^{jk}n_{k}),
\q
where $P_{i j}=\delta_{i j}-n_{i} n_{j}$ is the projection operator that realizes the transversal gauge in the $\bm{n}$-direction. According to \cite{Landau:1982dva}, the total energy emitted due to the electromagnetic dipole and quadrupole radiations is
\ee
\frac{dE_{EM}}{dt}=-\frac{r^2}{4\pi} \int \bm{H}^ 2 d\Omega
\q
where $H^i=\epsilon^i_{jk}\dot{A}^{j} n^k$. Adding Eq.~\eqref{A} and using
\ee
\int    n^{i} d \Omega=\int    n^{i} n^{j} n^{k}d \Omega=0,
\q
\ee
\int  n^{i} n^{j} n^{k} n^{l} d \Omega=\frac{4 \pi}{15}\left(\delta^{i j} \delta^{k l}+\delta^{i k} \delta^{j l}+\delta^{i l} \delta^{j k}\right), \quad \int  n^{d} n^{k} d \Omega=\frac{4}{3} \pi \delta^{d k},
\q
we find
\ee
\frac{dE_{EM}}{dt}=\frac{dE_{EM}^{dip}}{dt}+\frac{dE_{EM}^{quad}}{dt},
\q
where
\ee
\frac{dE_{EM}^{dip}}{dt}=-\frac{2\ddot{p}^2}{3},
\q
\ee
\frac{dE_{EM}^{quad}}{dt}=-\frac{\ddddot{D}_{ij}\ddddot{D}_{ij}}{20}.
\q
Similarly, the loss rate of angular momentum due to electromagnetic radiation is given by
\ee
\frac{d L_{EM}^{i}}{d t}=- \frac{r^2}{4\pi}\int  (-\epsilon^{i k l}\left(\partial_{0} A_{j}\right) x^{k} \partial^{l} A_{j}+\epsilon^{i k l} A_{k} \partial_{0} A_{l}) d \Omega.
\q
After very careful calculation, we obtain
\ee
\frac{dL_{EM}^{i}}{dt}=\frac{dL_{EM}^{i,dip}}{dt}+\frac{dL_{EM}^{i,quad}}{dt},
\q
where
\ee
\frac{dL_{EM}^{i,dip}}{dt}= -\frac{2\epsilon^{i k l}}{3}\dot{p}_k \ddot{p}_l
\q
\ee
\frac{dL_{EM}^{i,quad}}{dt}=-\frac{1 }{20 } \epsilon^{i k l}\left(\ddot{D}_{k a} \dddot{D}_{l a}\right)
\q
Noting
\ee
\frac{dE_{GW}^{quad}}{dt} \equiv  -\frac{1}{5}\left(\dddot{Q}_{ij}\dddot{Q}_{ij}\right), \quad \frac{d L_{GW}^{i,quad}}{d t}\equiv -\frac{2 }{5 } \epsilon^{i k l}\left(\ddot{Q}_{k a} \dddot{Q}_{l a}\right)
\q
and using the definition of traceless charge quadrupole \eqref{definition}, we find the relations between quadrupole radiations:
\ee
\frac{dE_{EM}^{quad}}{dt}\equiv\frac{\mu^{2}\left(Q_{1} / m^2_{1}+Q_{2}/ m^2_{2}\right)^{2}}{4} \frac{dE_{GW}^{quad}}{dt},
\q
\ee
\frac{dL_{EM}^{i,quad}}{dt}\equiv \frac{\mu^{2}\left(Q_{1} / m^2_{1}+Q_{2}/ m^2_{2}\right)^{2}}{8}  \frac{dL_{GW}^{i,quad}}{dt}.
\q
Under the condition of gravitational attraction dominance over the electric repulsion, i.e $|Q_1 |\leq m_1$ and $| Q_2|\leq m_2$, we always have $\frac{dE_{EM}^{quad}}{dt}\leq \frac{1}{4} \frac{dE_{GW}^{quad}}{dt}$ and $\frac{dL_{EM}^{i,quad}}{dt} \leq \frac{1}{8} \frac{dL_{GW}^{i,quad}}{dt}$.

\bibliographystyle{JHEP}
\bibliography{Ref}

\end{document}